\documentclass[twocolumn]{aastex701}

\usepackage{amsmath}
\usepackage{booktabs}

\usepackage{lineno}
\linenumbers

\begin{document}

\title{
Beyond Mergers: the Role of Metallicity in Forming a Reservoir of Massive \\ and Detached Binary Black Hole Systems}

\author[0009-0007-7564-4408]{Siyuan Chen}
\affiliation{Center for Astrophysics $\vert$ Harvard \& Smithsonian, 60 Garden Street, Cambridge, MA 02138, USA}
\email{siyuan.chen@cfa.harvard.edu}  

\author[0000-0003-0648-2402]{Monica Gallegos-Garcia}
\affiliation{Center for Astrophysics $\vert$ Harvard \& Smithsonian, 60 Garden Street, Cambridge, MA 02138, USA}
\affiliation{Harvard Society of Fellows, 78 Mount Auburn Street, Cambridge, MA 02138, USA}
\email{monica.gallegos_garcia@cfa.harvard.edu}

\begin{abstract}
Isolated binary evolution is a key pathway to forming binary black hole (BBH) mergers. However, the observed gravitational-wave population indicates it may not account for all mergers, especially those with high primary black hole mass ($\gtrsim 30 M_{\odot}$). 
Using the binary population synthesis code \texttt{POSYDON}, we simulate populations of binary systems at eight metallicities from $10^{-4}$ to $2\,Z_\odot$ and classify each BH-forming binary as a merging BBH, a detached BBH, or a system disrupted during BH formation. 
We find that metallicity shapes these populations differently, especially for high BH masses ($\gtrsim 25 M_{\odot}$). 
At lower metallicities, between $10^{-2}$ and $0.45\,Z_\odot$, massive BHs continue to form but become increasingly unlikely to merge: the number of massive merging BBHs declines sharply, while the detached population remains abundant. 
For $25$--$35\,M_\odot$ BHs, the detached-to-merging ratio increases by more than two orders of magnitude across this metallicity range. 
After weighting by cosmic star formation, massive BHs are found predominantly in detached binaries with orbital periods  $\gtrsim 10^2$ days. 
Therefore, isolated binary evolution can form massive BBHs even if it rarely produces massive BBH mergers, leaving behind a massive and detached population that may be driven to merger through dynamical interactions in dense stellar environments.

\end{abstract}

\section{Introduction}
The catalog of gravitational-wave (GW) detections has made it possible to study a growing population of merging binary black holes \citep[BBHs;][]{gwtc-3, LVK_O4_2025, gwtc4pop}. 
The diversity of observed BBH properties suggests that multiple formation channels contribute to the population \citep[e.g.,][]{Zevin2021, Colloms2025}, and subpopulations of these mergers become increasingly associated with different formation pathways \citep[e.g.,][]{LIGOScientific:2026ctl,Ray_subpop_2026,Cheng:2026bpc,Galaudage:2026opk,Alvarez-Lopez:2026ymo,Flanagan:2026ayy,Flanagan:2026btu,Guttman2026,Ray:2026qer, 2026arXiv261007152K}.
These subpopulations can help distinguish both where a proposed formation channel is compatible with the observations as well as where it is not.
In particular, the subpopulations of the most massive BBHs ($m_1 \gtrsim 30 M_{\odot}$) appear less naturally explained by isolated binary evolution than lower-mass systems \citep[e.g.,][]{Banagiri2025, Sridhar2025, Guttman:2026cnv, Ray_subpop_2026, Galaudage:2026opk}.  
To fully understand the origin of these systems, it is necessary to establish not only which binaries a given channel can bring to merger, but also the systems that it does not.

Isolated binary evolution is one of the most widely studied pathways for producing tight compact-object binaries that can merge within a Hubble time \citep[for recent reviews, see][]{Mapelli2021, MandelBroekgaarden2022, MandelFarmer2022}.
For isolated binary evolution, even though the BBH merger efficiency decreases by several orders of magnitude toward higher metallicity \citep[e.g.][]{Giacobbo2018,GiacobboMapelli2018,Klencki2018,Belczynski2010_met,Neijssel2019,Broekgaarden2022,EldridgeStanway2016,2025ApJ...979..209V}, some studies \citep[e.g.][]{Giacobbo2018,Iorio2023} suggest that the total BBH formation efficiency depends more weakly on metallicity.
Metallicity therefore affects the fate of BBH systems more than whether a BBHs is formed at all. 

The non-merging population provides a variety of different systems. 
BHs in detached binaries with stellar companions are observed with Gaia \citep{2023MNRAS.518.1057E, 2023MNRAS.521.4323E, 2024A&A...686L...2G}, wide BBHs may be detectable through microlensing \citep[e.g.][]{Eilbott2017,Abrams2025}, and detached BBHs in dense stellar environments may be hardened through dynamical encounters and driven to merger \citep[e.g.,][]{Rodriguez:2016vmx, 2020ApJS..247...48K}. 

In this paper, we examine how metallicity determines whether BBH progenitors ultimately merge, remain detached, or become disrupted, and how the relative importance of these pathways varies with BH mass.
Quantifying these populations is necessary both for understanding the observed BBH merger population and for identifying detached systems that may be detected through other observational probes or serve as a reservoir for other formation pathways, such as dynamical hardening in dense stellar environments.

The occurrence of the various outcomes depends on the interactions the progenitor systems experience, and thus on if and how the progenitor stars expand. 
This dependence is especially important for massive systems. 
Most population synthesis studies rely on rapid codes based on analytic fits to single-star models, which extrapolate the expansion of the most massive stars beyond the range of the models from which the fits were constructed and do not self-consistently include stellar winds \citep{Hurley2000}. 
Several studies have shown that these fits overestimate the maximum radii of massive stars compared to more recent stellar-evolution models \citep[e.g.][]{2026arXiv260821609G,2026arXiv260628506M}.
This is the case especially for more massive stars   where 1D stellar evolution codes tend to find that they can remain compact, especially towards near-solar metallicity \citep[e.g.][]{2023MNRAS.525..706R, 2023NatAs...7.1090B, 2024ApJ...964L..23R, 2024A&A...692A.141K}. 

Using detailed models \texttt{MESA} \citep{Paxton2011,Paxton_2013, Paxton2015, Paxton_2018, Paxton_2019, Jermyn2023} integrated into \texttt{STARTRACK} \citep{2002ApJ...572..407B, 2008ApJS..174..223B} with \texttt{METISSE} \citep{2020MNRAS.497.4549A, 2023MNRAS.525..933A}, \citet{2023MNRAS.525..706R} find that differences in radial expansion may have little effect on lower-mass BBH mergers, for which Roche-lobe overflow often occurs before the donor reaches its maximum radius,
However, for the most massive systems ($M_\mathrm{tot} \gtrsim50~M_\odot$), the treatment of stellar expansion can substantially change the binary outcome: models with limited expansion reduce the predicted merger rate by factors of $\sim 15$--$50$\footnote{These results may be underestimated, since the integration of detailed stellar models in standard rapid binary population synthesis codes still causes an overestimate of stellar radii in binary systems due to ``track jumping’’ \citep[see Section in 3.2][]{2026arXiv260821609G}.}. 
Nevertheless, while the absolute radii remain uncertain, recent detailed calculations generally find that the most massive stars expand less than previously calculated. 

To follow massive binaries from the zero-age main sequence (ZAMS) through BH formation with detailed models, we use \texttt{POSYDON} \citep{Fragos2023, Andrews2025}, which uses grids of detailed \texttt{MESA} binary simulations.
\citet{Kiato2026} used \texttt{POSYDON} to characterize various properties of the Galactic population of BHs in single and binary configurations.
Among their results, they found a peak in the mass distribution of detached systems and provided calculations for microlensing events and {\it Gaia} BHs. 
Here, we focus on the metallicity dependence of three outcomes involving only BHs (merging BBHs, detached BBHs, and BHs from disrupted systems) across $10^{-4}$ to $2\,Z_\odot$. 
We separate our analysis across BHs mass ranges to show how metallicity plays different roles in determining the evolutionary outcome of low and high-mass BHs in binary systems. 

This paper is structured as follows. In Section~\ref{sec:methods} we describe our \texttt{POSYDON} simulations. In Section~\ref{sec:results} we present the merging, detached, and disrupted populations across metallicity, the stellar and binary evolution that produces them, and their cosmological rates. We discuss caveats and compare to previous work in Section~\ref{sec:discussion}, and summarize our conclusions in Section~\ref{sec:conclusion}.

\section{Method}
\label{sec:methods}

We use the binary population synthesis code \texttt{POSYDON} v2.1.0 \citep{Fragos_2023, Andrews2025} to model the evolution of binary systems. Here, we summarize the physical assumptions most relevant to the formation of BBHs\footnote{
The \texttt{POSYDON} configuration used in this work, together with the population files required to reproduce the principal figures, calculations, and results can be found at \dataset[https://zenodo.org/records/23169255]{https://zenodo.org/records/23169255}.  
The  POSYDON v2.0 grids used in this work are available on Zenodo
under data release v2.0 
\dataset[doi.org/10.5281/zenodo.15194708]{https://zenodo.org/records/21909570}; \citet{doi.org/10.5281/zenodo.15194708}.
}.

\texttt{POSYDON} simulations are based on grids of single- and binary-star models computed with the 1D stellar-evolution code \texttt{MESA} \citep{Paxton2011,Paxton_2013, Paxton2015, Paxton_2018, Paxton_2019, Jermyn2023}. 
These grids span eight metallicities, $Z/Z_\odot=10^{-4}$, $10^{-3}$, $10^{-2}$, 0.1, 0.2, 0.45, 1.0 and 2.0, adopting $Z_{\odot}=0.0142$ \citep{2009ARA&A..47..481A}. 
At each metallicity, we sample $10^6$ binaries. 
Primary masses are sampled from the \citet{2001MNRAS.322..231K} initial mass function (IMF) over $7-150~M_\odot$, while the mass ratio $q=M_2/M_1$ is drawn from a uniform distribution for $M_2 \geq 0.5~M_\odot$. 
Initial orbital periods are drawn from the \citet{2012Sci...337..444S} distribution over $0.75-6000$ days, and all binaries are initialized on circular orbits. 
We assume a binary fraction of unity.

\subsection{Stellar Physics}
We briefly summarize the wind and compact-remnant prescriptions most relevant to BH formation.
For hot, hydrogen-rich stars $>8M_{\odot}$, \texttt{POSYDON} adopts the wind mass-loss prescription of \citet{2000A&A...362..295V} (hereafter line-driven Vink winds), with a metallicity dependence of $(Z/Z_\odot)^{0.68}$. 
For hot, helium-rich stars, with surface hydrogen mass fraction below 0.4, Wolf-Rayet-like winds are modeled following \citet{2000A&A...360..227N} (hereafter WR winds).
\texttt{POSYDON} incorporates an LBV-type wind prescription: 
stars reaching the Humphreys-Davidson limit \citep[HD;][]{1979ApJ...232..409H}  [$L>6\times10^5~L_\odot$ and $(L/L_\odot)^{1/2} \times (R/R_\odot) > 10^5$)] are given an enhanced mass-loss of $10^{-4}\ M_{\odot}~{\mathrm{yr}}^{-1}$ following \citet{ 2010ApJ...715L.138B}.
Convective energy transport is treated with mixing-length theory (MLT) with $\alpha_\mathrm{MLT}=1.93$, while MLT++ prescription of \texttt{MESA} is applied in radiation-dominated, superadiabatic regions to improve numerical convergence \citep{Paxton_2013, Fragos_2023}. 
For a subset of stripped-helium-star models, \texttt{POSYDON} v2 further modifies the MLT++ activation thresholds to facilitate convergence during envelope expansion \citep[Sec. 4.1 in ][]{Andrews2025}. See Sec.~\ref{sec:caveats} for additional discussion on MLT++.

For BH formation, we explore four core-collapse prescriptions: the \citet{Sukhbold_2016} and \citet{2020MNRAS.499.2803P} prescriptions, each evaluated using the N20 and W20 engines.  
The mass lost to neutrinos during core collapse is capped at $0.5~M_\odot$. Pulsational pair-instability and pair-instability SNe are treated using the modified \citet{2023MNRAS.526.4130H} prescription implemented in \texttt{POSYDON}, adopting $\Delta M_{\rm CO-core}=0$ and $\Delta M_{\rm PPI}=-20~M_\odot$ \citep{Andrews2025}.
Natal-kick magnitudes are drawn from a Maxwellian distribution with a one-dimensional dispersion of $\sigma=265~\mathrm{km~s^{-1}}$ \citep{2005MNRAS.360..974H}. We adopt the mass-scaled prescription $v_{\rm kick,BH} = v_{\rm draw} \times 1.4M_\odot/M_{\rm BH}$, where $v_{\rm draw}$ is drawn from the underlying Maxwellian distribution.
The magnitude of BH natal kicks remains uncertain, with observations spanning substantial kicks \citep[e.g.,][]{Willems2005,Fragos2009,Kimball2023} as well as small or negligible ones \citep[e.g.,][]{Banagiri2023,Gomez2024,Burdge2024}.
\textit{Gaia} DR3 kinematics similarly support a mixed population in which some BHs receive strong kicks and others essentially none \citep{Nagarajan2025}. 
We therefore adopt mass-scaled kicks, which naturally suppress kicks for the most massive BHs while retaining substantial kicks at lower masses.

\subsection{Binary Evolution Physics}
We now summarize the treatment of mass transfer (MT), MT stability, and common-envelope (CE) evolution.
For MT between two non-degenerate stars, \texttt{POSYDON} uses the \texttt{contact} scheme in \texttt{MESA} for Roche-lobe overflow initiated on the main sequence. This allows both stars to fill their Roche lobes simultaneously and form a contact system \citep{2016A&A...588A..50M}. 
For post-main-sequence donors, MT is calculated using the \texttt{kolb} scheme \citep{1990A&A...236..385K}. 

Accretion onto a BH is limited to the Eddington rate. For a stellar accretor, the specific angular momentum carried by the accreted material is calculated following \citet{2013ApJ...764..166D}, which can rapidly spin the accretor to critical rotation. \texttt{POSYDON} adopts rotation-limited accretion: as the accretor approaches critical rotation, enhanced winds remove mass and angular momentum, preventing the star from rotating above the critical limit. Because tides can spin down the accretor and thereby allow further accretion, the MT efficiency is not prescribed a priori but emerges from the coupled evolution of accretion, rotation, winds, and tides.

MT stability is determined self-consistently within each detailed \texttt{MESA} binary model. MT is classified as unstable if the MT rate exceeds $0.1\ M_{\odot}~{\mathrm{yr}}^{-1}$ or if material overflows an outer Lagrange point. For a more detailed discussion of the uncertainties associated with the latter condition, see Appendix A of \citet{2026arXiv260901784G}. For systems with a BH accretor, instability is also assumed when the photon-trapping radius exceeds the Roche-lobe radius of the BH. 
Contact systems are classified as unstable if either component has evolved beyond the main sequence. 
These criteria identify systems expected to undergo runaway MT or rapid orbital inspiral owing to the loss of material with high specific orbital angular momentum.

When any of these instability criteria is reached, the detailed \texttt{MESA} calculation is terminated and the subsequent CE evolution is treated beginning with the standard $\alpha-\lambda$ energy formalism \citep[for reviews, see][]{Ivanova_2013,2023pbse}.  
We define the envelope boundary at a hydrogen mass fraction of $X_{\rm H}=0.3$ for hydrogen-rich donors and a helium mass fraction of $Y_{\rm He}=0.1$ for stripped helium-star donors.
The envelope-binding parameter $\lambda$ is obtained from the corresponding detailed \texttt{MESA} models at the onset of CE. 
We adopt the optimistic prescription for Hertzsprung-gap donors, allowing them to survive CE evolution when the envelope can be successfully ejected. Following successful envelope ejection, the layers between the adopted envelope boundary and the core (defined by $X_\mathrm{H}=0.01$) are removed through an instantaneous, fully non-conservative SMT phase.
Our $\alpha_{\rm CE}=1$ simulation suite includes all four core-collapse prescriptions described above. 
To assess the sensitivity of our results to the assumed CE efficiency, we additionally calculate populations with $\alpha_{\rm CE}=0.5$ and 2.0 for the \citet{2020MNRAS.499.2803P}+W20 and \citet{Sukhbold_2016}+N20 prescriptions which can be found in Appendix~\ref{sec:appendix_alphaCE}.

\subsection{Cosmological Rates Calculation} \label{method:cosmo_rate}

To estimate the rates of the merging, detached, and disrupted BBH populations, we use the rate calculation implemented in \texttt{POSYDON} \citep[see Section 6.3 of][]{Andrews2025}.
Each synthetic binary is assigned a statistical weight, in units of $\mathrm{yr}^{-1}$, based on the cosmic star-formation and metallicity histories of the IllustrisTNG-100 simulation \citep{2018MNRAS.480.5113M, 2018MNRAS.477.1206N, 2018MNRAS.475..648P, 2018MNRAS.475..676S, 2019ComAC...6....2N}.
These weights are converted into volumetric rate densities by dividing by the comoving volume associated with each redshift bin.
For the analysis, we evaluate rate densities to $z\le2$, an observationally motivated but optimistic upper bound on the current and near-term LVK reach for massive BBHs \citep{2026ApJ..1005L..51A}. For merging BBHs, the redshift cut is applied at the time of merger. For detached and disrupted systems, it is instead applied at the formation of the second BH, when the system either remains bound as a wide, detached BBH or becomes unbound during the second core collapse. 
We do not consider BHs formed through stellar mergers.

\section{Results}
\label{sec:results}

\subsection{Evolutionary Outcomes Across Metallicities}\label{sec:evolutionary_outcomes_per_z}
\label{subsec:evolution_metallicity}

\begin{figure}[t!]
\centering
\includegraphics[width=0.48\textwidth]{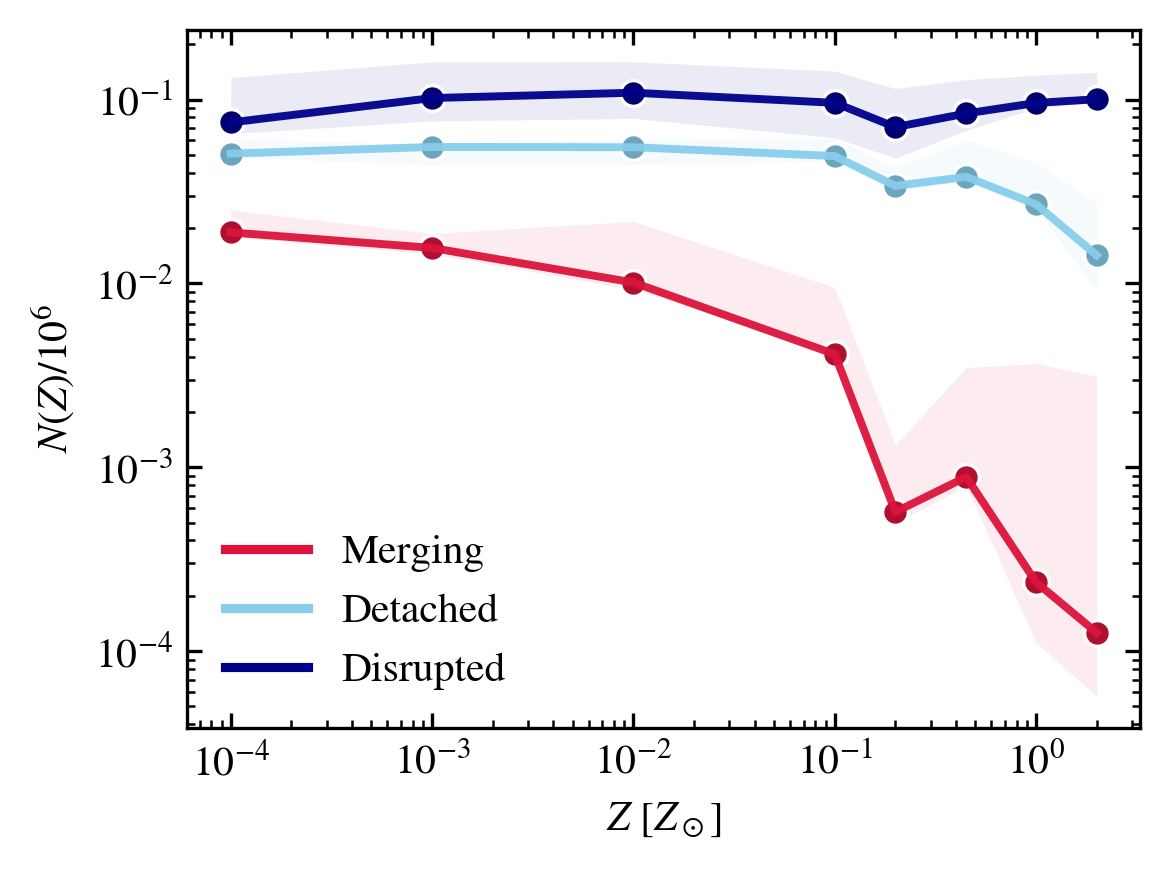}
\caption{Fraction of the $10^6$ binaries simulated at each metallicity that produce merging (red), detached (light blue), or disrupted BH systems (blue). Solid lines show the results obtained using the \cite{2020MNRAS.499.2803P} supernova prescription with the \texttt{W20} engine. Shaded regions encompass results obtained using the same prescription with the \texttt{N20} engine and the \citet{Sukhbold_2016} prescription with either the \texttt{W20} or \texttt{N20} engine.}
\label{fig:total_num}
\end{figure}

We first present the global trends of each evolutionary outcome with metallicity before showing how BHs of different masses contribute to each population.

In Fig.~\ref{fig:total_num}, we show the fraction of binaries from our $10^6$ populations per metallicity that produce merging, detached, and disrupted BH systems. 
Merging BBHs are subdominant to both detached and disrupted systems across the full metallicity range, and their fraction declines steeply above $0.1~Z_\odot$. 
The detached fraction also decreases toward high metallicity, whereas the disrupted fraction remains approximately constant. 
Though not to the same degree, the evolutionary outcomes of binaries in our simulations depend strongly on metallicity.

These trends are consistent with previous population-synthesis studies, which find that the BBH merger efficiency decreases toward higher metallicity across a range of codes and physical assumptions \citep[e.g.,][]{Belczynski2010_met, EldridgeStanway2016, Giacobbo2018, GiacobboMapelli2018, Klencki2018, Neijssel2019, Broekgaarden2022, Kiato2026}. 
In our models, stronger line-driven winds at higher metallicity remove more mass, widen the orbit, and produce less massive stellar cores and BHs. 
Under our mass-scaled kick prescription, the resulting lower-mass BHs receive larger natal kicks, while the wider orbits are less tightly bound, making these systems more likely to become disrupted (see the $2\,Z_\odot$ example in Fig.~\ref{fig:evolution}). 
Systems that remain bound have longer GW inspiral times because of their wider orbits and lower BH masses, shifting systems that would otherwise merge into the detached population.
Metallicity-dependent stellar expansion provides an additional effect \citep[e.g.,][]{Iorio2023, 2023MNRAS.525..706R}. 
In rapid population-synthesis models, the larger radii predicted at higher metallicity can cause binaries to merge as stars rather than form BBHs \citep{2025ApJ...979..209V}. 
We do not include stellar mergers in our populations, but this channel may further reduce the number of BBHs at high metallicity.
The nearly constant disrupted fraction reflects two competing effects: the fraction of BH-forming binaries that are disrupted increases with metallicity, from $\approxeq 52\%$ at $10^{-4}\,Z_\odot$ to $\approxeq 88\%$ at $2\,Z_\odot$ (Tables~\ref{tab:mass_bins_metallicity_S1} and \ref{tab:mass_bins_metallicity_S2}), but this is offset by a decrease in the total number of binaries that form BHs at high metallicity.

\begin{figure*}[t!]
\centering
\includegraphics[width=1\textwidth]{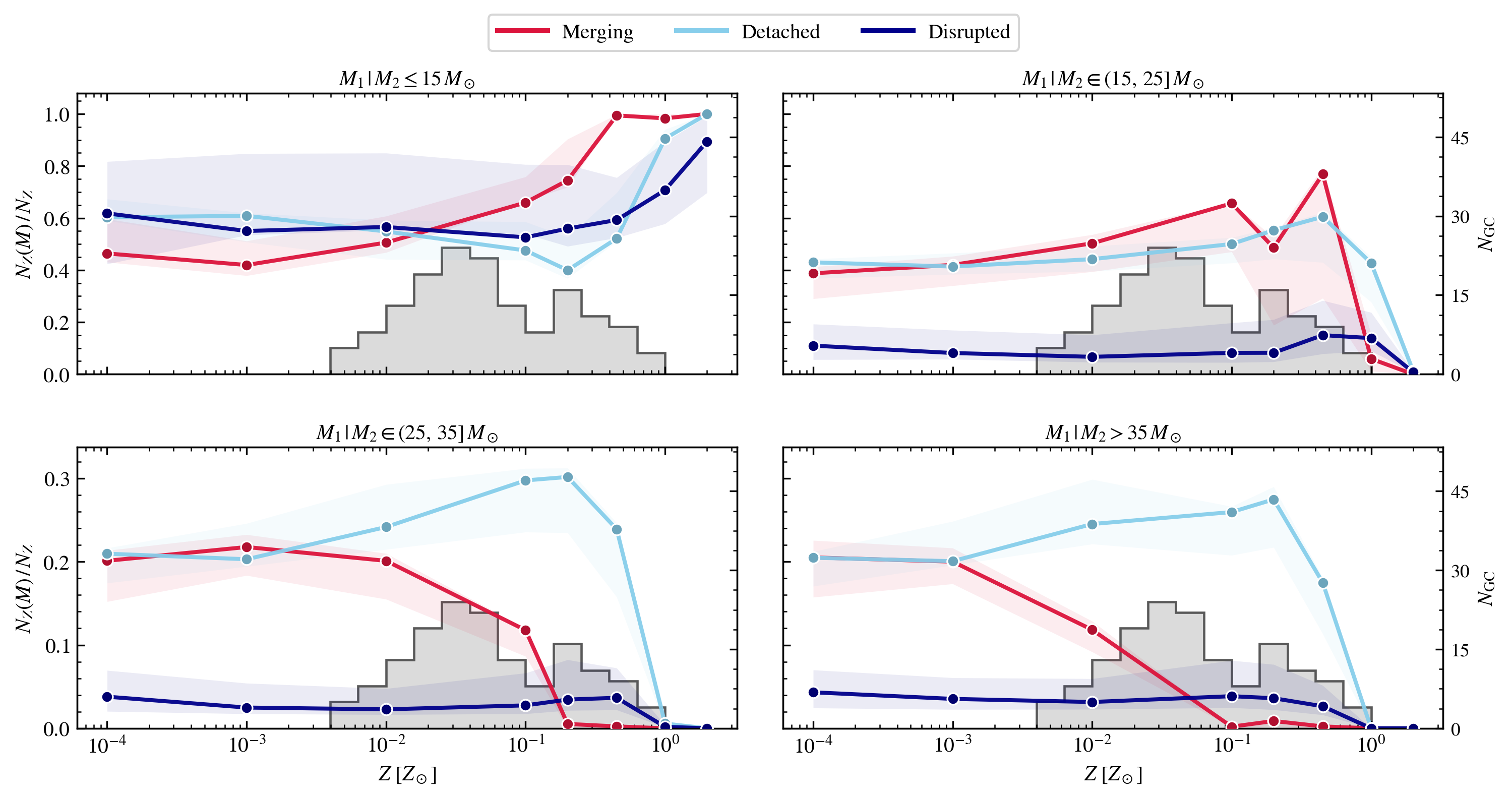}
\caption{Fractional contribution of binaries containing at least one BH in the indicated mass range for each category as a function of metallicity. Each fraction is normalized by the total number of binaries at the corresponding metallicity. The panels represent different BH mass ranges, while the colors distinguish merging, detached, and disrupted populations. Solid lines show results obtained using the \citet{2020MNRAS.499.2803P} supernova prescription with the \texttt{W20} engine. Shaded regions span the results obtained with the corresponding \texttt{N20}-engine model and with the \citet{Sukhbold_2016} prescription using the \texttt{W20} or \texttt{N20} engine. The grey histograms show the metallicity distribution function for the globular clusters in the Milky Way.}
\label{fig:total_num_mZ}
\end{figure*}

We next examine how BHs of different masses contribute to each evolutionary outcome. 
Fig.~\ref{fig:total_num_mZ} shows the fraction of systems containing at least one BH with a given mass interval for the merging (red), detached (light blue), and disrupted (dark blue) populations.
The corresponding absolute numbers for the number of primary or secondary BHs in each mass interval and metallicity are listed separately in Tables~\ref{tab:mass_bins_metallicity_S1} and \ref{tab:mass_bins_metallicity_S2} in the Appendix~\ref{sec:appendix_num}. 
The absolute counts follow the same overall metallicity trends as the system-level fractions shown in Fig.~\ref{fig:total_num_mZ} which we describe below.
In Fig.~\ref{fig:total_num_mZ} we also overlay the Galactic globular cluster (GC) metallicity distribution \citep{1959MNRAS.119..538K, 1985ApJ...293..424Z} for discussion in Sec.~\ref{subsec:reservoir_gc}. 
The GC metallicities are binned uniformly in [Fe/H] at 0.2~dex intervals, using bin edges defined on a grid anchored at [Fe/H] = 0. The bin edges are then mapped to total metallicity assuming $Z/Z_\odot = 10^{[\mathrm{Fe/H}]}$.
This approximation assumes a solar-scaled abundance pattern and may underestimate the total metallicity at low $[\mathrm{Fe/H}]$, where $\alpha$-element enhancement is common due to the delayed iron enrichment from Type Ia supernovae relative to core-collapse supernovae.

The top left panel in Fig.~\ref{fig:total_num_mZ} shows the contribution of low-mass BHs ($M_\mathrm{BH} \le 15~M_\odot$) to each outcome as a function of metallicity.  
These low-mass BHs account for an increasing fraction of all three populations with  increasing metallicity, as more massive BHs become rarer overall (see Tables~\ref{tab:mass_bins_metallicity_S1} and \ref{tab:mass_bins_metallicity_S2} for detailed counts).
Binary systems with BHs within the $15-25~M_\odot$ range (top right panel) similarly become more abundant up to intermediate metallicities before declining at the highest $Z$, where 
most BHs have masses below $15~M_\odot$ (see absolute numbers in the Appendix~\ref{sec:appendix_num}). 
For these two mass ranges, we find that all three contributions to the population have a similar dependence with metallicity. 

\begin{figure*}
    \centering
    \includegraphics[width=1\linewidth]{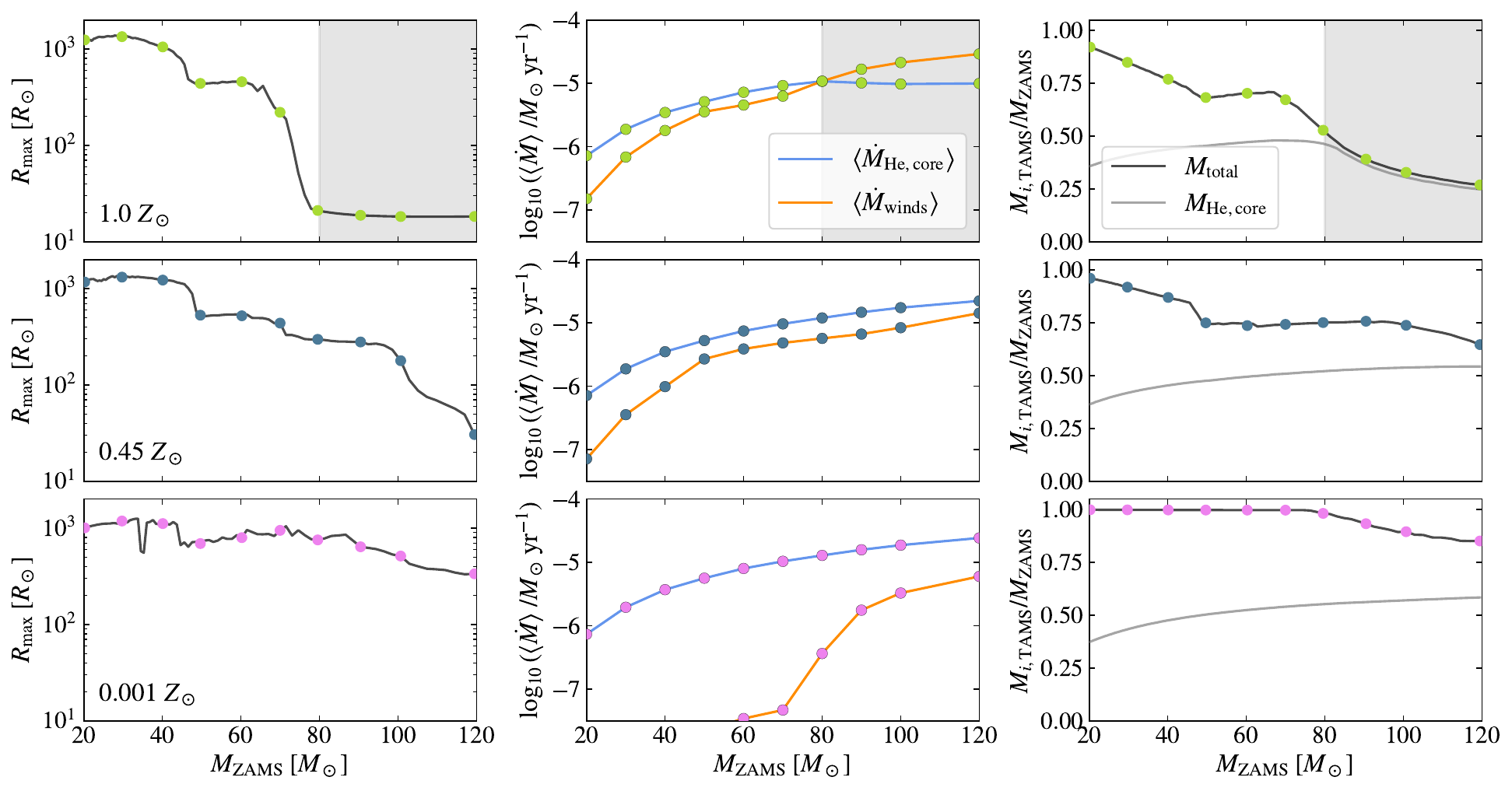}
    \caption{Properties of single-star models as a function of ZAMS mass $M_{\mathrm{ZAMS}}$, at solar metallicity (top row), $0.45,Z_{\odot}$ (middle row), and $0.001,Z_{\odot}$ (middle row). 
    Left: maximum radius $R_{\mathrm{max}}$ reached over the star's evolution. 
    Middle: average mass-loss rate due to stellar winds during the MS, $\langle\dot{M}_{\mathrm{winds}}\rangle$, compared to the average growth rate of the helium core during the MS, $\langle\dot{M}_{\mathrm{He,core}}\rangle$. 
    Right: mass at TAMS normalized by $M_{\mathrm{ZAMS}}$, showing the total mass $M_{\mathrm{total}}$ (black) and the helium core mass $M_{\mathrm{He,core}}$ (gray); the gap between the two curves is the remaining envelope mass. 
    The shaded region in the top row marks where $\langle\dot{M}_{\mathrm{winds}}\rangle$ exceeds $\langle\dot{M}_{\mathrm{He,core}}\rangle$.
    } 
    \label{fig:single_star_MS_evolution}
\end{figure*}

If metallicity affected BH progenitors of all masses in the same way, every mass bin would show the same metallicity dependence.
Although this is the case for binary systems below $25M_{\odot}$, this breaks down at higher masses. 
In the $25-35~M_\odot$ range (bottom left panel), the contribution of these BHs to the merging fraction begins to decline by $0.01~Z_\odot$ and becomes negligible by $0.2~Z_\odot$, whereas the detached fraction continues to rise to $0.45~Z_\odot$ and drops at $1~Z_\odot$. 
The absolute number of binaries follows this trend: between $0.01$ and $0.45~Z_\odot$, the numbers of primary and secondary BHs within this mass range in merging systems decrease from 1337 to 2 and from 1611 to 0, respectively, while the detached population remains of order $10^4$ up to $0.45~Z_\odot$ for both primary and secondary BHs. 
In other words, the ratio between the number of detached massive BHs and merging BHs increases with metallicity from 11.3 to 3325 between $0.01$ and $0.45~Z_\odot$. 
The divergence in trend with metallicity is even stronger for higher-mass BH ($> 35M_{\odot}$):  by $0.1~Z_\odot$ there are only eight total merging BHs and less at higher metallicities, while 7771 detached BHs remain at $0.45~Z_\odot$. 
Therefore, the fractional and absolute populations, as well as the detached-to-merging ratio, show that metallicity does not affect all BH masses equally. 
For massive BHs, the merging channel is suppressed at substantially lower metallicity than the detached channel. 
This results in a population of massive and detached BBH systems.

\subsection{Single Star Evolution at Varying Metallicities}\label{sec:single_binary_met}

In the previous section, we showed that the evolutionary outcomes of low- and high-mass BBH progenitors depend strongly on metallicity. 
Here, we examine two aspects of stellar evolution that provide insights into these trends: wind-driven mass loss and post-main-sequence expansion. 
We first describe how the maximum radii of single stars vary with initial mass and metallicity. 
We then consider a representative massive binary system with fixed ZAMS component masses and initial orbital period to demonstrate how metallicity alter its evolutionary outcome.

In this section, our goal is not to provide an exhaustive explanation for the detailed behavior of the stellar models or the exact maximum radii they reach. 
Instead, we use these results to illustrate how metallicity-dependent wind mass loss and stellar radii can lead to a spectrum of binary interactions and outcomes. 

In Fig.~\ref{fig:single_star_MS_evolution} we show maximum radius $R_{\mathrm{max}}$ (left panel), average mass loss via winds $\langle\dot{M}_\mathrm{winds}\rangle$ and average core growth rates during the main-sequence evolution $\langle\dot{M}_\mathrm{He,core}\rangle$ (middle panel), and interior mass distribution at TAMS (right panel), shown as a function of ZAMS mass $M_{\mathrm{ZAMS}}$ for stars at solar (top row), $0.45 Z_{\odot}$ (middle row), and $0.001 Z_{\odot}$ (bottom row). 

At solar metallicity, $R_{\mathrm{max}}$ ranges from $\simeq 20$ to $10^{3}~R_{\odot}$. 
We find that this range is strongly influenced by the envelope mass remaining at terminal-age main sequence (TAMS), while $R_{\mathrm{max}}$ is also sensitive to the MLT prescription, as discussed in Sec.~\ref{sec:caveats}.
In the middle panel, we show that $\langle\dot{M}_\mathrm{winds}\rangle$ is comparable to $\langle\dot{M}_\mathrm{He,core}\rangle$ across this mass range, and that $\langle\dot{M}_\mathrm{winds}\rangle~>~\langle\dot{M}_\mathrm{He,core}\rangle$ for the most massive stars, with $M_{\mathrm{ZAMS}}\gtrsim 80\,M_{\odot}$. 
This is because the line-driven winds during the main sequence increase with metallicity and reach mass-loss rates of $\approx 10^{-5}\,M_{\odot}\,\mathrm{yr}^{-1}$ for the most massive stars in this range \citep{2026ApJ...997...52C}, resulting in a high $\langle\dot{M}_\mathrm{winds}\rangle$.
The gray region in the top row of Fig.~\ref{fig:single_star_MS_evolution} marks where $\langle\dot{M}_\mathrm{winds}\rangle > \langle\dot{M}_\mathrm{He,core}\rangle$. 
In this region, wind mass loss dominates over core growth during the MS, substantially depleting the stellar envelope by TAMS (right panel). 
With little envelope mass remaining, these stars reach only small $R_{\mathrm{max}}$. 
This behavior is consistent with previous studies using detailed stellar models \cite[e.g.][]{2023NatAs...7.1090B, 2024ApJ...964L..23R, 2024A&A...692A.141K}, which also show that strong winds at high metallicity can strip the envelope and suppress stellar expansion.
Although several wind prescriptions are used during the full evolution of these stars, strong line-driven Vink winds during the main sequence alone are sufficient to substantially suppress the stellar radius, even without including LBV winds \citep[see Fig.~2 in][]{2023NatAs...7.1090B}.
We refer the reader to Fig. 3 in \citet{Andrews2025} for a full accounting of mass loss wind schemes used across both $M_{\mathrm{ZAMS}}$ and four representative metallicities in \texttt{POSYDON}. 
Stars outside the gray region 
can also experience substantial mass loss during the MS. 
These stars lose up to $\simeq 0.25\,M_{\mathrm{ZAMS}}$ but retain $\simeq 0.2\,M_{\mathrm{ZAMS}}$ in their envelopes. 
With a larger fraction of envelope material, these stars expand to larger radii.

At $0.45\,Z_{\odot}$ (middle row), a similar trend appears, except that stars with $M_{\mathrm{ZAMS}}\gtrsim 80\,M_{\odot}$ do not experience dominant $\langle\dot{M}_\mathrm{winds}\rangle $ strengths and retain up to $\approxeq 0.2 M_{\mathrm{ZAMS}}$ in envelope material at TAMS.
These stars now reach $R_{\mathrm{max}}\simeq 200$--$400\,R_{\odot}$. 
At the lowest metallicity (bottom row), the envelopes at TAMS are substantially more massive than those at higher metallicities. 
These stars experience relatively little mass removal during the MS, except at the largest masses ($\gtrsim 90M_{\odot}$) where LBV winds occur during the main sequence.
The larger envelopes likely allow these stars to expand farther. 
Consequently, stars with $M_{\mathrm{ZAMS}}\lesssim 80\,M_{\odot}$ reach $R_{\mathrm{max}}$ values of up to $\simeq 10^{3}\,R_{\odot}$ at low metallicity. 

Several factors beyond the remaining envelope mass influence $R_{\mathrm{max}}$, including how metallicity affects at what evolutionary stage a star expands \citep[e.g.][]{2020A&A...638A..55K}. 
We therefore do not attempt to explain the exact maximum radius of every model. 
Instead, these comparisons across metallicity show that metallicity-dependent wind mass loss plays a strong role in setting the available envelope mass and, in turn, the overall scale of post-main-sequence expansion.

\begin{figure}[t!]
\centering
\includegraphics[width=0.49\textwidth]{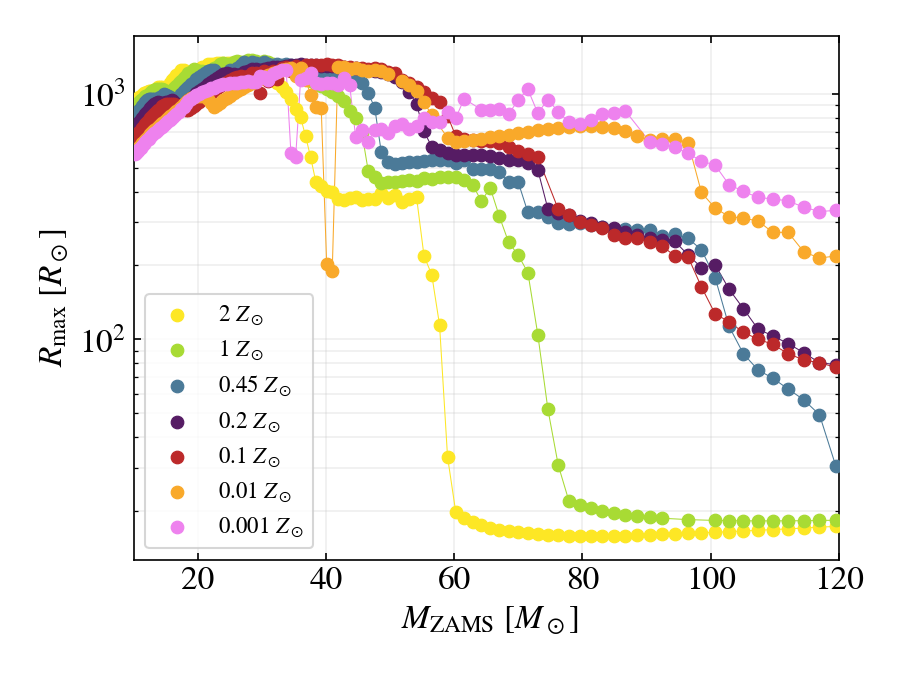}
\caption{Maximum stellar radius as a function of ZAMS for seven metallicities.  }
\label{fig:radius_ZAMS}
\end{figure}

In Fig.~\ref{fig:radius_ZAMS} we overlay $R_{\mathrm{max}}$ for several metallicities to summarize two trends.
First, for stars with $M_{\mathrm{ZAMS}}~\gtrsim~40\,M_{\odot}$, $R_{\mathrm{max}}$ decreases with increasing metallicity, potentially limiting the range of possible binary interactions.  
This behavior is opposite to that of lower-mass stars, for which $R_{\mathrm{max}}$ increases with metallicity. 
Second, the variation in $R_{\mathrm{max}}$ across $M_{\mathrm{ZAMS}}$ becomes larger with increasing metallicity: at lower metallicity, low- and high-mass stars reach more similar maximum radii compared to high metallicity where the maximum radii across the mass spectrum are more distinct. 
For high-mass stars, the maximum radius varies substantially more with metallicity, leading to 
more variation in evolutionary outcomes in massive stars. 

This strong metallicity dependence is broadly consistent with a set of detailed stellar models of \citet{2023MNRAS.525..706R} (Model 4b using \texttt{MESA} without MLT++ shown in Fig. 2). 
They show that for stars with $M_{\mathrm{ZAMS}} \lesssim 40\,M_{\odot}$, the maximum radius generally increases with metallicity. 
However, for larger masses and higher metallicities, at $Z \simeq 0.015$ (approximately solar metallicity under our adopted value of $Z_{\odot}$), a $50\,M_{\odot}$ star reaches a smaller maximum radius than a $40\,M_{\odot}$ at the same metallicity.
Similar to Fig.~\ref{fig:radius_ZAMS}, this effect becomes stronger at higher metallicities: at $Z \simeq 0.02~Z$, their $50\,M_{\odot}$ model reaches a smaller maximum radius than even a $20\,M_{\odot}$ model at the same metallicity.
This result is found both with and without the MLT++ treatment, suggesting that it is not specific to this modeling choice; we discuss this further in Section~\ref{sec:caveats}.

We note that the models in \citet{2023MNRAS.525..706R} reach larger maximum radii than those in this work. 
For the $50\,M_{\odot}$ model, they find $R_{\mathrm{max}} \approxeq 1000\,R_{\odot}$, compared to $\simeq 450\,R_{\odot}$ in our models. 
However, the overall trend is the same: at higher metallicity, the maximum radius of massive stars is smaller than their low-mass counterparts.
The trend with metallicity described here is in contrast to the single-star models used in rapid binary population synthesis codes which show that the maximum radii increases with ZAMS mass at high metallicity, $\gtrsim 0.1 Z_{\odot}$, see Fig. 1 in \citet{2023MNRAS.525..706R} and Fig. 8 in \citet{2025ApJ...979..209V} for examples.

\subsection{An Representative Example of Binary Star Evolution at Varying Metallicities}\label{sec:only_binary_met}

\begin{figure*}[t!]
\centering
\includegraphics[width=\textwidth]{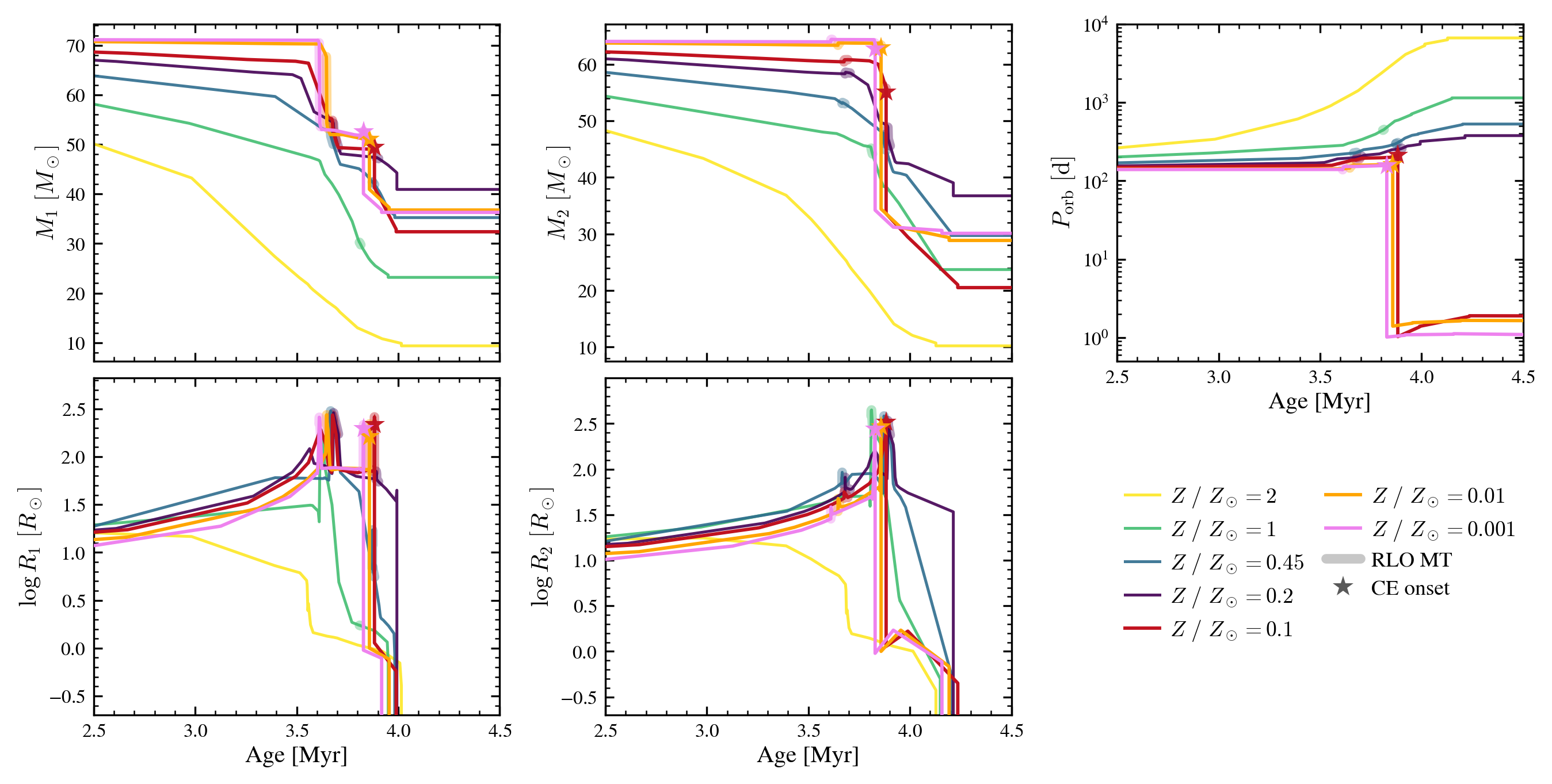}
\caption{Evolution of identical binary systems at different metallicities. The binary systems are initially $72.2~M_\odot+66.6~M_\odot$ with an orbital period of 125 days. We do not apply a natal kick to create identical conditions across binary systems. The panels show the evolution of the component masses, radii, and orbital period. The three lowest-metallicity systems lead to BBH mergers, while all others remain detached. When natal kicks are included, the highest-metallicity binary is disrupted.}
\label{fig:evolution}
\end{figure*}

Given these trends in maximum stellar radius and impact of strong stellar winds with metallicity, in Fig.~\ref{fig:evolution} we compare six binaries with the same ZAMS component masses and initial orbital period but different metallicities. 
This comparison shows how changing only the metallicity can shift the evolutionary outcome of massive systems from a BBH merger to a detached BBH.
Natal kicks are not included here to enable a more consistent comparison by minimizing random effects.

For the binary systems at $0.001$ (pink), $0.01$ (orange) and $0.1~Z_\odot$ (red), the primary fills its Roche lobe after entering core-helium burning, at an age of $\simeq3.7$~Myr and a radius of $\simeq200~R_\odot$.
This initiates the first MT episode at only $\simeq20\%$ of the maximum radius reached by the corresponding star in isolation (Fig.~\ref{fig:radius_ZAMS}). 
This MT phase remains stable and reverses the mass ratio without removing all hydrogen from the primary’s envelope. 
The now more massive secondary subsequently leaves the main sequence and initiates MT at $\simeq 3.9$~Myr. 
This second episode becomes dynamically unstable as both stars evolve off the main sequence, triggering a double CE evolution that reduces the orbital period from $\approxeq 200$ days to $\approxeq 1$ day. 
The resulting BBHs merge within a Hubble time.

For binary systems at $Z=0.2~Z_\odot$ (purple) and $0.45~Z_\odot$ (blue), the same post-main-sequence MT episodes remain stable.
In this case, the binaries continue through MT without entering a CE phase, ultimately forming detached BBHs with orbital periods of several hundred days. 
At $Z=1~Z_\odot$ (green) only the secondary undergoes MT, while neither star fills its Roche lobe at $Z=2~Z_\odot$ (yellow). 

The different outcomes of the second MT phase arise from several metallicity-dependent effects on stellar structure and its impact on the binary configuration. 
The structure of the initially more massive star at the onset of the second MT phase is particularly important. 
At solar metallicity, it has already been stripped to a compact helium star with $R_1\simeq1.7~R_\odot$. 
At $Z=0.45~Z_\odot$, it retains only a thin H-rich envelope and remains relatively compact, with $R_1\simeq7~R_\odot$, whereas at $Z=0.2~Z_\odot$ less stripping leaves a more extended primary with an H-rich envelope of $\simeq 6~M_\odot$ and $R_1\simeq 60~R_\odot$. 
This shows that with decreasing metallicity, the initially more massive star retains more mass and is larger at the onset of the second MT phase. 
During the second MT, the initially less massive star expands only modestly to $\sim 70~R_\odot$ and remains well within its Roche lobe. 
The evolution of the initially less massive star and the orbit also contributes to the different outcomes after the second MT: at $0.2$ and $0.45~Z_\odot$, the initially less massive star is larger after the first MT episode than in the lower-metallicity models and has crossed the LBV wind criterion, leading to enhanced mass loss (as shown in the $M_2$ panel) before the onset of the second MT. 
This additional stripping can limit its subsequent expansion during the second MT since the enhanced stripping contracts the star. 
At the same time, the higher-metallicity ($\gtrsim 0.2~Z_{\odot}$) systems enter the second MT phase at wider separations, providing larger Roche lobes and further favoring stable MT.
At $Z=0.1~Z_\odot$, the similarly extended primary undergoes sustained expansion during the second MT, reaching $R_1\simeq260~R_\odot$ comparable to its Roche-lobe radius. The primary therefore also fills its Roche lobe, driving the binary into contact and subsequently into CE evolution. 
The transitions through the evolutionary outcomes show that metallicity impacts multiple aspects of binary interaction history. 
The general trend shown here is that at higher metallicity, stronger wind mass loss and differences in stellar radial evolution make Roche-lobe overflow less likely, consistent with the maximum-radius comparison above. 
For the systems that undergo two MT episodes, the second interaction begins at wider separations and under different stellar conditions, and is more often stable in our simulations.
In particular, our results are consistent with \citet{Kiato2026} who also show a trend of fewer interactions with increasing metallicity for BBH binaries from massive stellar progenitors (see their Fig. 11).

The remnant masses exhibit a non-monotonic dependence on metallicity. At high metallicity, stronger line-driven winds continuously remove mass producing progressively lower-mass BHs. 
Reducing the metallicity weakens the winds and allows the stars to retain more mass, with the largest final BH masses obtained at $0.2~Z_\odot$. 
However, the final masses decrease towards lowest metallicities because both stars undergo substantial post-main-sequence expansion and envelope stripping. 
In this example, the final BH masses are set by competing effects of wind mass loss and binary interactions: weaker winds towards lower metallicity can increase the mass available to form a BH, but greater expansion towards lower metallicities also leads to more MT and CE evolution which strips the envelopes and reduces the final BH mass.
The final BH masses therefore depend non-monotonically on metallicity.

In these systems we did not include natal kicks. 
Incorporating natal kicks leaves the mass and radius evolution nearly unchanged. 
The outcomes also remain the same, except for the $Z=2~Z_\odot$ system, which is disrupted upon formation of the second BH. 
This higher-metallicity system is particularly susceptible to disruption because its wide orbit makes the binary weakly bound, and  its low BH masses result in larger kicks by the adopted mass-scaled kick prescription.
Although this comparison represents only one initial binary configuration, it cleanly illustrates how metallicity shapes the properties and ultimate fate of the system, especially whether it forms a sufficiently tight BBH to merge within a Hubble time.

\subsection{Astrophysical Population}\label{sec:astro_pop}
\begin{figure*}[t!]
\centering
\includegraphics[width=0.9\textwidth]{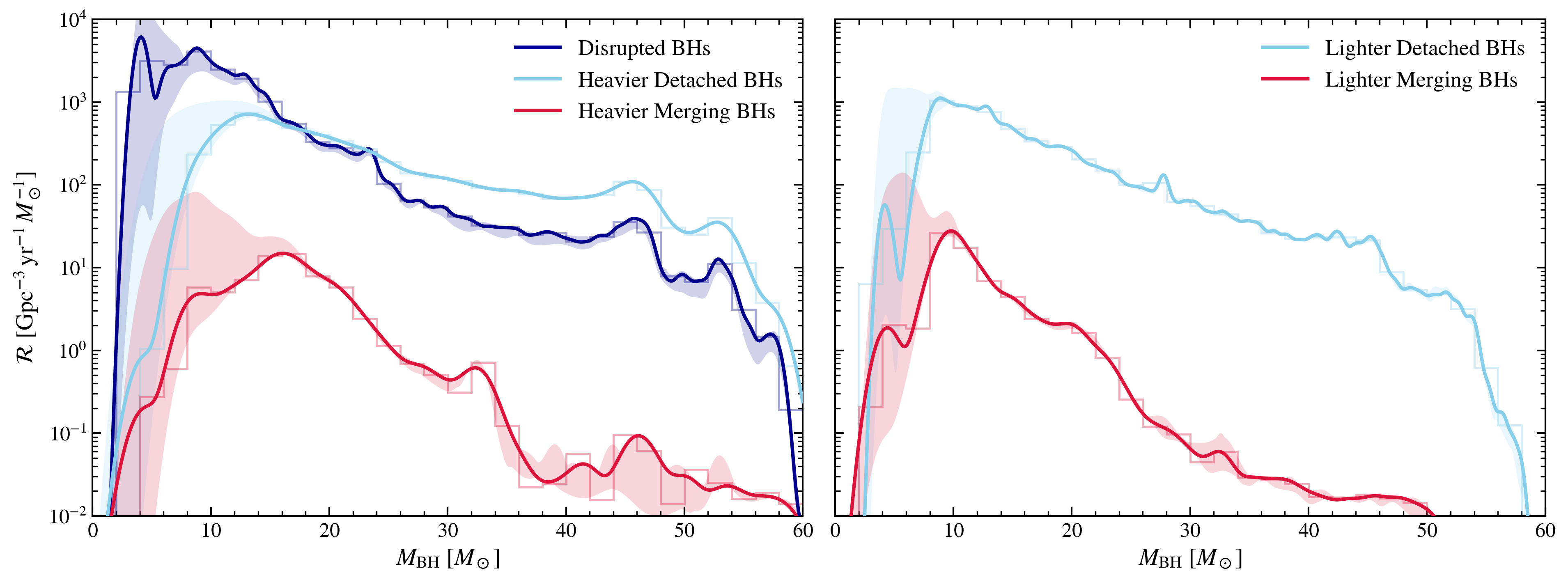}
\caption{Rate density as a function of BH mass for different populations within redshifts $z<2$. The left panel shows the heavier black hole in each merging or detached binary, together with all black holes from disrupted binaries. The right panel shows the lighter black hole in each merging or detached binary. Disrupted black holes are counted individually because the binary components are no longer bound. The bands span the regions for different supernova prescriptions.}
\label{fig:mass_BH}
\end{figure*}

In this section, we now focus on the astrophysical BH population at $z<2$, where the metallicity-dependent evolutionary trends discussed above can be mapped onto the local population of merging, detached, and disrupted systems. 
Weighting the simulated binaries by the cosmic star-formation history, metallicity distribution, and evolutionary delay times (as discussed in Sec.~\ref{method:cosmo_rate}) allows us to assess how these different evolutionary channels contribute to the BH population in the low-$z$ Universe.

In Fig.~\ref{fig:mass_BH}, we show the rate densities as a function of BH mass for merging, detached, and disrupted astrophysical populations. 
We show these distributions for the more heavier BH in the BBH system (left) and for the lighter BH (right).
We highlight three results.
First, consistent with the population fractions in Fig.~\ref{fig:total_num}, merging BBHs are subdominant across the entire mass range, with rate densities about two orders of magnitude below those of detached and disrupted systems. 
Second, the detached population shows a shallower decline toward high BH masses than either the merging or disrupted population; in other words, as BH mass increases, BHs are predominantly found in detached binaries rather than in merging or a result of a disrupted systems.
These trends are consistent for both panels of Fig.~\ref{fig:mass_BH}, i.e. for both BH components.
Third, while the disrupted systems dominate at low masses, for the heavier BHs, the detached population overtakes them at $\approxeq 20~M_\odot$ and remains dominant over the higher-mass regime. 
Overall we find that for higher-mass BHs ($\gtrsim 30~M_\odot$), the number of merging systems decreases sharply, while detached BBHs remain numerous, consistent with the trend seen in Fig.~\ref{fig:total_num_mZ}.

The shaded bands span the rate densities predicted by four SN prescriptions. 
The differences are largest at low BH masses ($\lesssim 15~M_\odot$) and weaken at higher masses. 
This convergence shows the similar treatment of massive progenitors where BH formation increasingly occurs through direct collapse \citep[e.g.][]{Sukhbold_2016,2020MNRAS.499.2803P}. 
The low-mass BH population is more sensitive to the differences in progenitor explodability, particularly the transition between successful explosions that produce NSs and failed explosions that form BHs \citep{Fragos_2023}. 
Despite these quantitative differences, the overall mass dependence remains consistent across all four SN models.

\section{Discussion}
\label{sec:discussion}
\subsection{A reservoir of massive and detached BBHs} \label{subsec:reservoir_gc}
\begin{figure}[t!]
\centering
\includegraphics[width=0.48\textwidth]{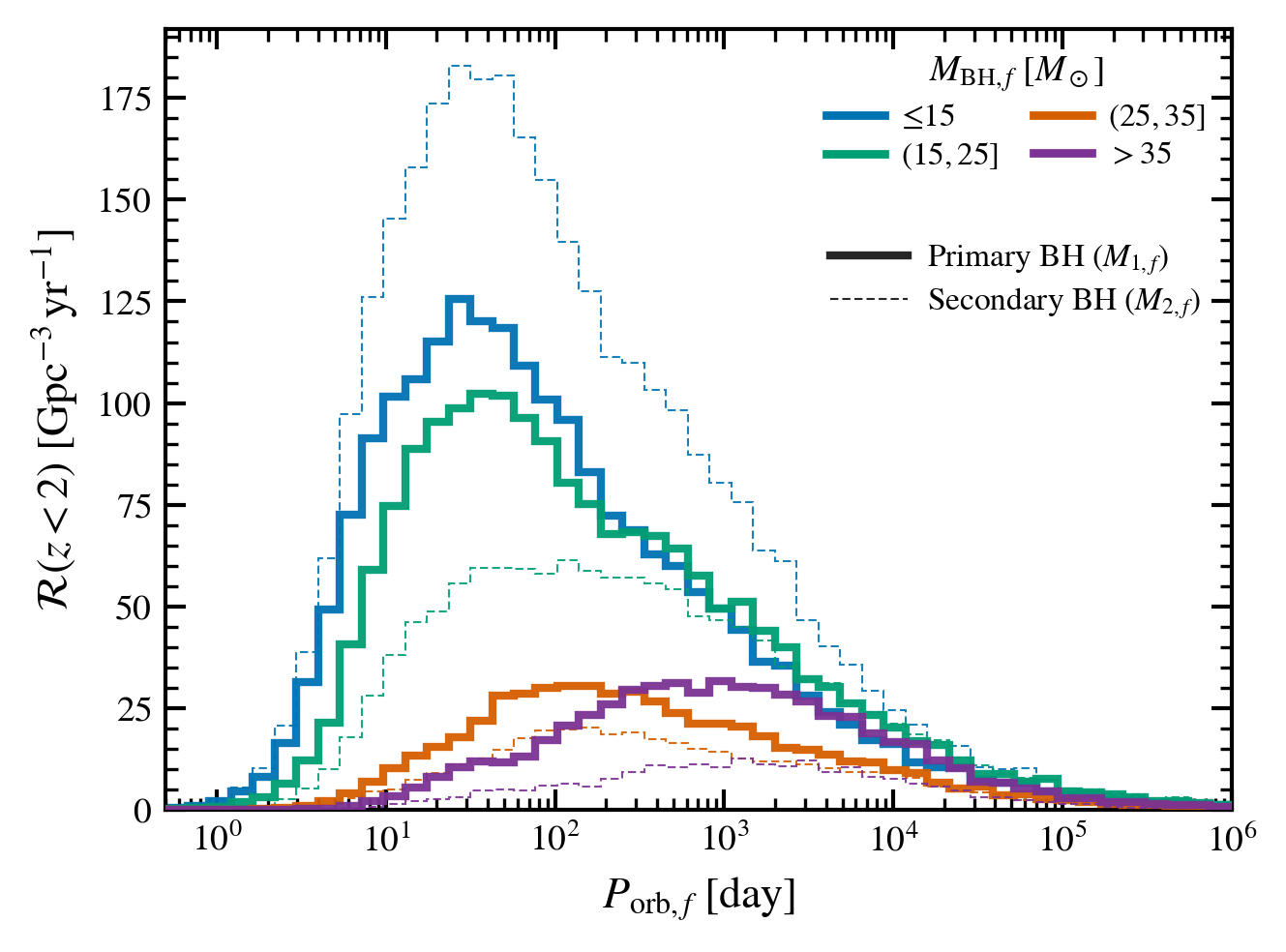}
\caption{Rate density of detached BBHs at $z<2$ as a function of final orbital period with colors showing the final BH mass ranges. Solid and dashed lines show the primary and secondary BHs respectively.}
\label{fig:final_orbit}
\end{figure}

In Sec.~\ref{subsec:evolution_metallicity} we identified a metallicity range 
$0.01 \lesssim Z/Z_\odot \lesssim 0.45$ in which massive ($\gtrsim 25~M_\odot$) detached BBHs remain abundant while the number of merging BBHs decreases sharply, causing the detached-to-merging ratio to increase by over two orders of magnitude. This metallicity range overlaps with that of the metal-rich subpopulation of globular clusters (Fig.~\ref{fig:total_num_mZ}).
Additionally, recent cluster models with refined metallicity coverage show that these metal-rich (``red'') clusters, $\mathrm{[Fe/H]}\gtrsim-1$, contribute a distinct population of dynamical BBH mergers \citep{Agrawal2026}.
In dense environments such as globular, nuclear, and open star clusters, detached BBHs serve as a population for future mergers: dynamical encounters can harden these binaries or increase their eccentricities, shortening their GW inspiral times \citep[e.g.,][]{2015PhRvL.115e1101R, 2016PhRvD..93h4029R, 2020ApJS..247...48K}.

To characterize the local detached BBH population in Section~\ref{sec:astro_pop}, we examine the orbital periods at the formation of the second BH.
In Fig.~\ref{fig:final_orbit}, we show the rate-density distributions of these orbital periods, which we separate into the different mass ranges. 
We find that the distributions shift systematically toward longer orbital periods with increasing BH mass.
The heavier (solid lines) and lighter (thin dashed lines) BHs follow the same trends. 
For $M_\mathrm{BH} \le 25~M_\odot$, the orbital periods peak around $\approxeq 30$ days while in the mass range of $25-35~M_\odot$, the orbital period peak near $\approxeq 10^2$ days. 
The highest mass BHs skew toward periods of thousands of days.
This trend likely reflects the combined effects of wind-driven orbital widening and binary survival: the more massive progenitors experience stronger wind mass loss, which tends to widen the orbit, while the weaker natal kicks associated with more massive BHs allow these wider systems to remain bound. In contrast, the stronger kicks received by lower-mass BHs preferentially disrupt wide binaries.
Thus, massive detached BBHs preferentially occupy wider orbits, even though wide binaries are more susceptible to disruption by natal kicks. 

The relevance of this population to GCs depends on several factors, such as the primordial binary fraction and the dynamical survival of these systems.
Present-day binary fractions in GCs \citep{2001ApJ...559.1060A, 2007MNRAS.380..781S, 2012A&A...540A..16M, 2015ApJ...807...32J, 2025A&A...693A.161M} are substantially lower than in the field \citep{2017ApJS..230...15M}. 
However, GCs may have formed with a field-like binary population, with the lower present-day fraction arising primarily from the dynamical disruption of soft binaries \citep[][see also \citealt{1992ApJ...389..527H, 2007ApJ...665..707H, 2009ApJ...707.1533F, 2015MNRAS.446..226L}]{2026arXiv260402412O}. 
More massive binaries remain hard at larger separations for a given velocity dispersion \citep{1975MNRAS.173..729H, 2026arXiv260402412O}. Thus, the massive detached BBHs in Fig.~\ref{fig:final_orbit}, with orbital periods of $\sim10^2$--$10^3$ days, may remain hard under typical GC conditions \citep{2018MNRAS.478.1520B, 2020ApJS..247...48K}. 
They could therefore survive as a population available for subsequent encounters that further harden the binaries and potentially drive them to merger \citep{1975MNRAS.173..729H, 2015ApJ...800....9M, 2015PhRvL.115e1101R, 2020ApJS..247...48K}.

Cluster models already include primordial binaries, typically evolved with rapid codes based 
on single-star fits. If massive stars remain compact, as in our models (Fig.~\ref{fig:radius_ZAMS}), 
fewer of these binaries would undergo CE evolution or stellar mergers, and more would instead 
remain as wide, detached BBHs. Primordial binaries could then supply dense clusters with a 
larger population of massive detached BBHs than currently predicted, strengthening a hybrid 
pathway in which isolated binary evolution forms the massive BBH and dynamical encounters 
subsequently drive it to merger. 
Such systems could contribute to the high-mass end of the 
cluster-origin merger population, which we will explore in a follow-up paper.

\subsection{Caveats}\label{sec:caveats}

The maximum radius reached by massive stars remains one of the major uncertainties in stellar-evolution models. This uncertainty directly affects our results: whether a massive binary interacts depends on how far its stars expand, while the changing internal structure during expansion affects whether that interaction is stable.

We first consider stellar winds. 
On the main sequence, \texttt{POSYDON} adopts the line-driven Vink winds for hot, hydrogen-rich stars. Recent theoretical and empirical studies suggest that the mass-loss rates of O-type stars may be lower by a factor of $\sim 3$ \citep[e.g.,][]{2014ARA&A..52..487S, 2022A&A...665A.133G}. 
Weaker main-sequence winds would allow stars to retain more of their envelopes and become larger and more luminous \citep{2022A&A...665A.133G}. 
However, lower main-sequence wind rates do not necessarily imply that the most massive stars reach large red-supergiant radii. 
For example, \citet{2024ApJ...964L..23R} adopt lower O-star mass-loss rates without MLT++, but their $60\,M_\odot$ solar-metallicity model still does not expand beyond $\sim 1000\,R_\odot$. 
Thus, although the absolute radii of our most massive models remain uncertain, massive stars may still remain relatively compact.

For stars that cross the Humphreys--Davidson limit, \texttt{POSYDON} applies an enhanced mass-loss rate of $10^{-4}\,M_\odot\,{\rm yr}^{-1}$ \citep{2010ApJ...714.1217B}. 
This prescription has little effect at solar metallicity in our models, where line-driven winds keep the most massive stars below the limit, but becomes increasingly important at lower metallicity \citep[e.g.,][]{2023NatAs...7.1090B, Andrews2025}.
Recent studies implement more physically motivated mass loss near the Eddington limit and find that it can similarly prevent the most massive stars from expanding to red-supergiant radii \citep{Cheng2024,2026A&A...707A..11P, 2026arXiv260728013S}. 
Although these studies differ from our prescription in both the mechanism and amount of mass loss, they support the physical plausibility of limiting the expansion of the most massive stars near the HD limit.

The treatment of convection in the superadiabatic, radiation-dominated layers of massive-star envelopes can also strongly affect their radii \citep[e.g.,][]{2012A&A...538A..40G, 2013ApJS..208....4P, 2015A&A...580A..20S, 2020A&A...638A..55K, 2022A&A...668A..90A, 2025A&A...693A.137R}. 
The MESA models in \texttt{POSYDON} use MLT++ \citep{Paxton_2013, Fragos_2023, Andrews2025}, which reduces the superadiabatic temperature gradient in these layers and limits the expansion of the most massive stars. 
Whether this treatment is appropriate remains uncertain \citep[see, e.g., Appendix A in][]{2020A&A...638A..55K} \citep[and Section 3.6 in][]{2025A&A...693A.137R, 2022A&A...668A..90A}. 
However, although MLT++ is a numerical prescription rather than a physical model, its effect resembles that of enhanced mass loss near the Eddington limit and may therefore produce reasonable radii for stars $\gtrsim 50\,M_\odot$. 
Supporting this interpretation, \citet{Cheng2024} model eruptive mass loss driven by locally super-Eddington layers and find that, for efficient mass loss, stars above $\sim20\,M_\odot$ do not evolve into red supergiants at either solar or SMC metallicity, consistent with the observed lack of red supergiants above $\log(L/L_\odot)\approx5.5$. 
The observational evidence remains mixed, however, with some comparisons favoring models with MLT++ and others favoring models without it \citep{2020A&A...638A..55K}.

\citet{2023MNRAS.525..706R} compare prescriptions affecting the maximum stellar radius using detailed MESA models computed both with and without MLT++. 
Their models without MLT++ also find that the most massive stars reach smaller maximum radii at higher metallicity, suggesting that this trend does not arise solely from the MLT++ prescription. 
These models recover the trend found here at near-solar metallicity, although they predict larger absolute radii (e.g., $R_{\rm max}\approx1250\,R_\odot$ compared with $\approx450\,R_\odot$ for a $50\,M_\odot$ model). 
Without MLT++, however, their most massive models encounter the numerical difficulties near the Eddington limit that MLT++ is designed to avoid, limiting this comparison to $M_{\rm ZAMS}\lesssim50\,M_\odot$. Removing MLT++ from our models would likely allow more of the most massive high-metallicity binaries to interact, but the overall metallicity trend would likely persist. 
These comparisons suggest that, although the absolute radii depend on the treatment of massive-star envelopes, limited expansion of the most massive stars at high metallicity is a common outcome across different approaches

\section{Conclusion}
\label{sec:conclusion}

We use \texttt{POSYDON} to evolve $10^6$ binaries at each of eight metallicities spanning $10^{-4}$--$2\,Z_\odot$. We classify each binary that forms a BH into one of three outcomes: a merging BBH, a detached BBH, or a system disrupted during compact-object formation.  
We further separating these outcomes by BH mass and find two distinct effects of metallicity on massive ($\gtrsim 25\,M_\odot$) BHs.

At $Z \gtrsim Z_\odot$, the formation of massive BHs is suppressed. Consequently, both the merging and detached populations become negligible in this mass range (Fig.~\ref{fig:total_num_mZ}; Tables~\ref{tab:mass_bins_metallicity_S1} and \ref{tab:mass_bins_metallicity_S2}). 
At lower metallicities, between $\sim 10^{-2}$ and $0.45\,Z_\odot$, massive BHs continue to form but become increasingly unlikely to merge. 
For $25$--$35\,M_\odot$ BHs, the number in merging systems decreases to nearly zero by $0.2\,Z_\odot$, while the detached population remains of order $10^4$ up to $0.45\,Z_\odot$.
 Over this range, the detached-to-merging ratio increases from 11.3 to 3325.

This divergence in trend is set by stellar and binary physics. In our models, massive stars expand less at higher metallicity (Figs.~\ref{fig:single_star_MS_evolution} and \ref{fig:radius_ZAMS}), causing CE evolution and stable MT to become less frequent. 
This behavior is consistent with other recent detailed stellar models, but differs from the single-star fits commonly used in rapid population synthesis.
The resulting binaries are therefore more likely to remain wide and detached rather than become BBH mergers (Fig.~\ref{fig:evolution}). 
When astrophysically weighted by cosmic star formation, we find that as BH mass increases  (especially $\gtrsim25M_{\odot}$), BHs are predominantly found in detached binaries rather than in merging or a result of a disrupted systems (Fig.~\ref{fig:mass_BH}), and these detached systems have orbital periods of $\gtrsim  10^2$ (Fig.~\ref{fig:final_orbit}).

Our results connect massive-star structure, binary interactions, and outcomes across BH populations. 
At intermediate metallicities, metallicity suppresses the formation of merging massive BBHs before it suppresses the formation of massive BHs themselves. 
Although the abundance of these systems remains sensitive to the uncertain expansion of massive stars, our results suggest that they may form a substantial detached population that, in suitable dynamical environments, can be hardened and driven to merger, potentially acquiring signatures of those interactions. 
These systems therefore warrant further investigation as potential dynamically-induced BBH mergers, as well as for targets of microlensing surveys.

\section*{acknowledgments}
The authors are grateful for helpful discussions with Enrico Ramirez-Ruiz, Morgan MacLeod, Max Briel, Elena Gonzalez and Edo Berger. 
The computations in this paper were run on the FASRC Cannon cluster supported by the FAS Division of Science Research Computing Group at Harvard University.
The POSYDON project is supported by the Gordon and Betty Moore Foundation (grant awards GBMF8477 and GBMF12341), the Swiss National Science Foundation (grant awards CRSII5\_213497 and 2000-1-243011), and the Foundation for the University of Geneva (FUNIGE; grant award SWISS-ET).

\textit{Software:} \texttt{POSYDON} \citep{Fragos_2023, Andrews2025}, \texttt{MESA} \citep{Paxton2011,Paxton_2013, Paxton2015, Paxton_2018, Paxton_2019, Jermyn2023}, \texttt{Matplotlib} \citep{2007CSE.....9...90H}, \texttt{NumPy} \citep{2020Natur.585..357H}, \textit{pandas} \citep{2010scpy.soft.....M}, \texttt{Astropy} \citep{2013A&A...558A..33A, 2018AJ....156..123A, 2022ApJ...935..167A}.

\appendix

\section{Absolute Number of Evolutionary Outcomes} \label{sec:appendix_num}

Here we provide the absolute numbers for BH systems shown in Fig.~\ref{fig:total_num_mZ} and discussed in Sec.~\ref{subsec:evolution_metallicity}.
As a reminder, for each metallicity, we evolve an independent population of $10^6$ binaries, with the initial stellar and orbital properties drawn from the distributions described in Sec.~\ref{sec:methods}, including the adopted IMF and distributions of binary parameters.
The values reported here are raw counts from these simulated populations and are not weighted by the cosmic star-formation or metallicity distribution.
We list both the absolute numbers of systems obtained from our population-synthesis simulations at each metallicity $N_{Z}$ and the number of BHs in each mass range $N_{Z}(M)$.  
We separate the systems into merging BBHs, detached BBHs, and disrupted with at least one BH and further distinguish according to the primary (Table~\ref{tab:mass_bins_metallicity_S1}) and secondary (Table~\ref{tab:mass_bins_metallicity_S2}) BH mass. 
For disrupted systems, the counts include binaries that form at least one BH before disruption and therefore do not necessarily correspond to BBHs.

\begin{table*}
\centering
\caption{Number of systems $N_Z(M)$ with $M_1$ in a given mass range, for the merging,
detached, and disrupted populations across metallicity. 
Masses are in $M_\odot$.
For the disrupted population, the counts refer to binary systems with at least one BH as not all systems are BBHs.}
\label{tab:mass_bins_metallicity_S1}
\begin{tabular}{ll rrrrrrrr}
\toprule
\multicolumn{2}{l}{$Z\ [Z_\odot]$} & 0.0001 & 0.001 & 0.01 & 0.1 & 0.2 & 0.45 & 1 & 2 \\
\midrule\midrule
Merging   & $N_Z$            & 18958 & 15611 & 10124 &  4126 &   572 &   886 &   237 &   126 \\
\cmidrule(lr){1-10}
          & $M \leq 15$      &  7554 &  5815 &  3616 &  1398 &   299 &   202 &   223 &   126 \\
          & $15 < M \leq 25$ &  5125 &  4794 &  4061 &  2285 &   265 &   680 &    14 &     0 \\
          & $25 < M \leq 35$ &  2688 &  2194 &  1337 &   440 &     3 &     2 &     0 &     0 \\
          & $M > 35$         &  3591 &  2808 &  1110 &     3 &     5 &     2 &     0 &     0 \\
\midrule\midrule
Detached  & $N_Z$            & 50882 & 55302 & 55119 & 49346 & 33943 & 38041 & 26934 & 14239 \\
\cmidrule(lr){1-10}
          & $M \leq 15$      & 17930 & 20691 & 16858 & 10577 &  5601 &  8656 & 17817 & 14180 \\
          & $15 < M \leq 25$ & 15173 & 16038 & 16167 & 16313 & 12846 & 17254 &  9060 &    59 \\
          & $25 < M \leq 35$ &  7790 &  7969 &  9150 & 10238 &  6721 &  6355 &    57 &     0 \\
          & $M > 35$         &  9989 & 10604 & 12944 & 12218 &  8775 &  5776 &     0 &     0 \\
\midrule\midrule
Disrupted & $N_Z$            & 75168 & 102047 & 109090 & 96126 & 70898 & 84119 & 95979 & 100535 \\
\cmidrule(lr){1-10}
          & $M \leq 15$      & 45576 &  71050 &  77818 & 63201 & 39636 & 44733 & 57770 &  71259 \\
          & $15 < M \leq 25$ &  7007 &   6820 &   5988 &  6674 &  4792 & 11601 & 11287 &    191 \\
          & $25 < M \leq 35$ &  2180 &   1891 &   1907 &  2000 &  1928 &  2832 &    60 &      0 \\
          & $M > 35$         &  1684 &   1949 &   1779 &  2322 &  1789 &  2085 &     0 &      0 \\

\bottomrule
\end{tabular}
\end{table*}

\begin{table*}
\centering
\caption{Same as Table~\ref{tab:mass_bins_metallicity_S1} but for $M_2$.}
\label{tab:mass_bins_metallicity_S2}
\begin{tabular}{ll rrrrrrrr}
\toprule
\multicolumn{2}{l}{$Z\ [Z_\odot]$} & 0.0001 & 0.001 & 0.01 & 0.1 & 0.2 & 0.45 & 1 & 2 \\
\midrule\midrule
Merging   & $N_Z$            & 18958 & 15611 & 10124 &  4126 &   572 &   886 &   237 &   126 \\
\cmidrule(lr){1-10}
          & $M \leq 15$      &  7714 &  5553 &  4428 &  2583 &   415 &   881 &   233 &   126 \\
          & $15 < M \leq 25$ &  5743 &  5191 &  3143 &  1398 &   156 &     5 &     4 &     0 \\
          & $25 < M \leq 35$ &  2592 &  2500 &  1611 &   140 &     1 &     0 &     0 &     0 \\
          & $M > 35$         &  2909 &  2367 &   942 &     5 &     0 &     0 &     0 &     0 \\
\midrule\midrule
Detached  & $N_Z$            & 50882 & 55302 & 55119 & 49346 & 33943 & 38041 & 26934 & 14239 \\
\cmidrule(lr){1-10}
          & $M \leq 15$      & 29935 & 32979 & 29523 & 22920 & 13366 & 19462 & 21873 & 14110 \\
          & $15 < M \leq 25$ & 12399 & 12904 & 13997 & 15144 & 12092 & 12454 &  4966 &   129 \\
          & $25 < M \leq 35$ &  4663 &  5141 &  6503 &  7021 &  5085 &  3930 &    95 &     0 \\
          & $M > 35$         &  3885 &  4278 &  5096 &  4261 &  3400 &  2195 &     0 &     0 \\
\midrule\midrule
Disrupted & $N_Z$            & 75168 & 102047 & 109090 & 96126 & 70898 & 84119 & 95979 & 100535 \\
\cmidrule(lr){1-10}
          & $M \leq 15$      & 28729 &  37680 &  40856 & 34742 & 34331 & 35836 & 48614 &  64196 \\
          & $15 < M \leq 25$ &  2021 &   2277 &   2073 &  1890 &  1887 &  2642 &  3842 &    612 \\
          & $25 < M \leq 35$ &   797 &    759 &    681 &   757 &   626 &   387 &   116 &      0 \\
          & $M > 35$         &  1631 &   1708 &   1712 &  1430 &   824 &   184 &     0 &      0 \\
\bottomrule
\end{tabular}
\end{table*}

\section{Impact of the Common-Envelope Efficiency Parameter $\alpha_{\rm CE}$} \label{sec:appendix_alphaCE}

We repeat our calculations shown in Fig.~\ref{fig:total_num_mZ} using the same initial conditions and physical assumptions but varying the CE efficiency parameter $\alpha_{\rm CE}$ to 0.5 and 2.0.
This parameter determines how efficiently orbital energy released during the inspiral is used to unbind the donor envelope: a larger $\alpha_{\rm CE}$ leads to more efficient envelope ejection and less orbital shrinkage, while a smaller $\alpha_{\rm CE}$ produces stronger orbital contraction and may result in merger during the CE phase.
Despite this dependence, we find that varying $\alpha_{\rm CE}$ has only a modest effect on the overall population statistics and on the BH mass distributions.

\begin{figure*}[t!]
\centering
\includegraphics[width=1\textwidth]{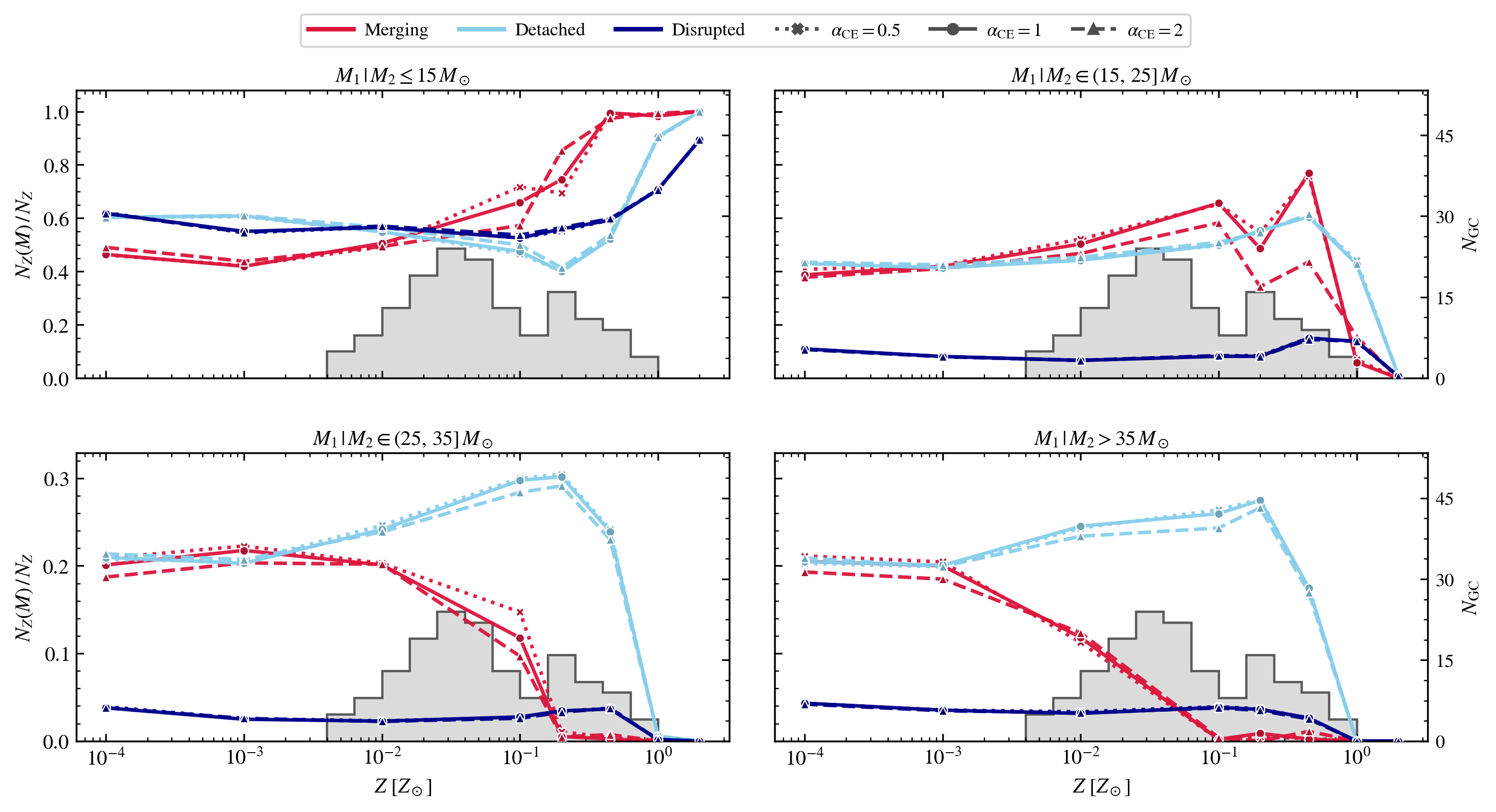}
\caption{Same as Fig.~\ref{fig:total_num_mZ}, but for different values of the CE efficiency parameter $\alpha_{\rm CE}$, using the supernova prescription of \citet{2020MNRAS.499.2803P} with the \texttt{W20} engine (corresponding to the solid lines in Fig.~\ref{fig:total_num_mZ}).}
\label{fig:alphas_total_mZ}
\end{figure*}

As shown in Fig.~\ref{fig:alphas_total_mZ}, the disrupted population is nearly insensitive to $\alpha_{\rm CE}$. Binaries that successfully emerge from CE evolution are typically left in relatively compact and tightly bound orbits, making them less susceptible to disruption by subsequent supernova mass loss and natal kicks. The abundance of detached BBHs is similarly stable across $\alpha_{\rm CE} \in [0.5,2]$, indicating that the channels contributing to this population are not strongly regulated by the CE efficiency. The merging population shows larger variations, as expected because its formation is more sensitive to the amount of orbital shrinkage during CE evolution: changing $\alpha_{\rm CE}$ changes whether a post-CE binary is sufficiently compact to merge within a Hubble time. Nevertheless, the qualitative metallicity dependence remains unchanged. In particular, over $0.01\lesssim Z/Z_\odot\lesssim0.45$, the number of massive BHs in merging systems declines sharply, while the massive detached population remains abundant. Our main conclusions regarding the emergence of a substantial detached-BBH reservoir at intermediate metallicities are therefore robust to variations in $\alpha_{\rm CE}$.

\bibliography{sample701, ms}{}
\bibliographystyle{aasjournalv7}
\end{document}